\documentclass[runningheads]{llncs}
\usepackage[T1]{fontenc}
\usepackage{graphicx,verbatim}
\usepackage{amsmath}
\usepackage{graphicx}
\usepackage{svg}
\usepackage{booktabs}
\usepackage{multirow}

\begin{document}
\title{Physics-Guided Radiotherapy Treatment Planning with Deep Learning}
%


\author{Stefanos Achlatis \inst{1,2} \and
Efstratios Gavves \inst{1} \and
Jan-Jakob Sonke \inst{1, 2}}
\authorrunning{S. Achlatis et al.}
%
\institute{
University of Amsterdam, Amsterdam, The Netherlands \and
The Netherlands Cancer Institute, Amsterdam, The Netherlands
\email{s.s.achlatis@uva.nl}\\}

\maketitle              

\begin{abstract}

Radiotherapy (RT) is a critical cancer treatment, with volumetric modulated arc therapy (VMAT) being a commonly used technique that enhances dose conformity by dynamically adjusting multileaf collimator (MLC) positions and monitor units (MU) throughout gantry rotation. Adaptive radiotherapy requires frequent modifications to treatment plans to account for anatomical variations, necessitating time-efficient solutions. Deep learning offers a promising solution to automate this process. To this end, we propose a two-stage, physics-guided deep learning pipeline for radiotherapy planning. In the first stage, our network is trained with direct supervision on treatment plan parameters, consisting of MLC and MU values. In the second stage, we incorporate an additional supervision signal derived from the predicted 3D dose distribution, integrating physics-based guidance into the training process. We train and evaluate our approach on 133 prostate cancer patients treated with a uniform 2-arc VMAT protocol delivering a dose of 62 Gy to the planning target volume (PTV). Our results demonstrate that the proposed approach, implemented using both 3D U-Net and UNETR architectures, consistently produces treatment plans that closely match clinical ground truths. Our method achieves a mean difference of 
\( D_{95\%} = 0.42 \pm 1.83 \) Gy and 
\( V_{95\%} = -0.22 \pm 1.87\% \) at the PTV while generating dose distributions that reduce radiation exposure to organs at risk. These findings highlight the potential of physics-guided deep learning in RT planning.


\end{abstract}

\section{Introduction}

Radiotherapy (RT) is a critical treatment modality for cancer, with approximately 50\% of cancer patients undergoing external beam radiation therapy during their disease~\cite{https://doi.org/10.1002/cncr.21324}. The aim of RT to deliver a sufficiently high radiation dose to the tumor while sparing surrounding healthy tissues as much as possible~\cite{Washington2020}. Over the past decades, various advanced delivery methods have been introduced to increase dose conformity and reduce toxicity. Notably, intensity-modulated radiation therapy (IMRT)~\cite{Bortfeld2006} and volumetric-modulated arc therapy (VMAT)~\cite{Otto2008} have transformed clinical practice. VMAT delivers highly conformal doses by continuously adjusting the leaves and jaws of the multileaf collimator (MLC) system while modulating the monitor units (MU) throughout the gantry rotation~\cite{Bedford_2009}.

Online adaptive radiotherapy (ART) personalizes RT by updating plans based on anatomical changes.~\cite{Lemus2024AdaptiveRadiotherapy}. This continual replanning process requires solving a high-dimensional, non-convex optimization problem~\cite{Unkelbach2015}. Although conventional algorithms based on Monte Carlo simulations~\cite{Verhaegen_2003_MonteCarlo} can be highly accurate, they are computationally expensive leading to inefficient workflows. To address these limitations, deep learning methods are increasingly being investigated as a way to accelerate treatment planning.

One prominent deep learning strategy is Deep Reinforcement Learning (DRL), where an agent learns to optimize treatment plans by maximizing a cumulative reward through repeated interactions with an environment~\cite{Sutton_Barto_2018}. DRL has shown promise in IMRT planning for prostate cancer~\cite{Shen_2021} and has been extended to VMAT planning~\cite{Hrinivich_Lee_2020,Hrinivich_2024}. In the latter case~\cite{Hrinivich_2024}, 3D collapsed cone convolution algorithms~\cite{Ahnesjo_1995_CollapsedCone} serve as the environment, while the Deep Deterministic Policy Gradient (DDPG) algorithm~\cite{lillicrap2019continuouscontroldeepreinforcement} is used for optimization. However, DRL-based approaches face several challenges, particularly their reliance on the accuracy of the environment and the design of reward functions used during training~\cite{reward_hacking}. 

An alternative strategy uses supervised deep learning on pre-calculated, clinically approved plans generated by commercial treatment planning systems (e.g., Pinnacle from Philips~\cite{Philips_Pinnacle} or Monaco from Elekta~\cite{Elekta_Monaco}). Early models used single-arc IMRT data and a four-layer 3D U-Net~\cite{cicek20163dunetlearningdense} to predict the MLC apertures from the patient’s Computed Tomography (CT) images and the corresponding masks of the Planning Treatment Volume (PTV) and organs-at-risk (OAR)~\cite{Ni_2022}. Later, an MU-decoder was added to predict both MLC configurations and MU values for three-arc breast cancer treatments as warm start for the optimisation~\cite{Vandewinckele_2023}. A key limitation of such direct supervision lies in the non-uniqueness of optimal solutions: multiple different MLC and MU configurations can produce clinically equivalent dose distributions, making a single ground truth plan inherently ambiguous.

To address these issues, we propose a two-stage, physics-guided \cite{faroughi2023physicsguidedphysicsinformedphysicsencodedneural}, \cite{pmlr-v250-vries24a} training pipeline for deep learning–based RT planning. In the first stage, our \emph{Deep RT Planner} is trained with direct supervision on the treatment plan parameters. In the second stage, we introduce an additional supervision signal derived from the 3D dose distribution corresponding to the predicted treatment plan, thereby incorporating physics guidance and training in a clinically relevant domain. This dose is generated by the \emph{RT Dose Predictor}, a fully differentiable gated recurrent unit (GRU) neural network~\cite{Cho_2014}, pretrained to predict 3D dose distributions from CT scans and treatment plans~\cite{Witte_Sonke_2024}, and remains frozen during the second stage of training.

We evaluate our method on a dataset of 133 patients with prostate cancer, all treated under a uniform 2-arc VMAT protocol delivering 62 Gy to the PTV. We implement two variants of our \emph{Deep RT Planner}: a 3D U-Net~\cite{cicek20163dunetlearningdense} with dual decoders—one for MLC masks and one for MU values, and a UNETR~\cite{Hatamizadeh_2022} architecture, which employs Vision Transformer (ViT)~\cite{Dosovitskiy_2021} encoders while retaining the dual-decoder structure. For both architectures, the second-stage physics-guided training significantly improves plan quality and dose accuracy.

\section{Method}

\subsection{Dataset}

We collected data from 133 prostate cancer patients treated at our institution between 2018 and 2022, following approval of the Institutional Review Board. For each patient, the planning CT scan and RT Structure Set (RTSS) were obtained, which includes the precise locations of the routinely considered CTV, PTV, rectum and femoral heads.

We recalculated all plans using Pinnacle~\cite{Philips_Pinnacle} with standardized parameters: 7 MeV beam energy, 20 treatment sessions, flattening filter-free mode, and a linear accelerator with 160 MLC leaves. Identical dose objectives were applied across all plans to ensure consistency, and an experienced radiation oncologist reviewed each case for clinical validity.

The final dataset included 104 training, 16 validation, and 13 test cases, each comprising a CT scan, binary masks for the CTV, PTV, and OARs, and a corresponding treatment plan. The CT volumes, centered on the isocenter, were resampled to a \(144 \times 144 \times 144\) grid (approximately \(500 \times 500 \times 500\) mm$^3$) with an isotropic resolution of \(3.5 \text{ mm}^3\). Hounsfield unit (HU) values were clipped to \([-1000, 3000]\) and normalized to \([-1, 1]\), with RTSS masks geometrically aligned to the CT grid.

Each VMAT plan consists of 144 control points (72 per arc). At each control point, a \(144 \times 144\) binary mask encodes the MLC aperture, representing MLC leaf and jaw positions relative to the isocenter. A corresponding scalar value specifies the monitor units (MU), resulting in 144 MLC aperture masks and 144 MU values per plan.

\subsection{Deep RT Planner and RT Dose Predictor}

For the \emph{Deep RT Planner}, we experimented with two architectures: a 3D U-Net~\cite{cicek20163dunetlearningdense} and a UNETR~\cite{Hatamizadeh_2022} model. Both architectures take as input a tensor of shape \( (B, C, D, H, W) \), where \( B \) is the batch size, $C$ is the number of channels, and D, H, and W are the depth, height, and widths. In our setting, $B=4$, $H=W=D=144$, and $C=5$. The first channel contains the CT scan, while the remaining four channels contain the rotation and projection at each control point for the CT, PTV, CTV, and OARs, respectively. This representation predefines the Beam's Eye View~\cite{Washington2020} at the input level, allowing the model to process spatially aware information. With a single forward pass, the models predict the complete RT plan, consisting of 144 MLC apertures and their corresponding monitor units.

\textbf{3D U-Net Architecture.} We employ a 3D U-Net with an encoder-decoder structure and skip connections across four resolution levels. The encoder applies repeated \( 3\times3\times3 \) convolutions with batch normalization and ReLU activation, followed by \( 2\times2\times2 \) max pooling. The decoder mirrors this structure, using upsampling before applying convolutional layers. Skip connections concatenate encoder features with decoder stages to retain spatial information. A final \( 1\times1\times1 \) convolution with sigmoid activation predicts the 144 binary MLC apertures.

A global average pooling layer extracts a latent representation from the deepest encoder features, which is processed by fully connected layers to predict the 144 MU values.


\textbf{UNETR Architecture.} We utilize a compact UNETR-based model with a lightweight ViT encoder and dual decoders for MLC mask and MU prediction. The 3D input volume is split into non-overlapping \( 16\times16\times16 \) patches, which are embedded into a latent space with positional encoding. A streamlined ViT with four transformer blocks (instead of twelve) processes the sequence, generating multi-scale feature representations while retaining intermediate states.

For MLC mask prediction, deconvolution and upsampling restore the original \( 144\times144\times144 \) resolution, with feature maps concatenated at multiple scales. A \( 1\times1\times1 \) convolution with softmax activation outputs the 144 binary MLC apertures. Meanwhile, the MU decoder branches from the ViT bottleneck and processes the latent sequence through fully connected layers to predict the 144 MU values.

\textbf{RT Dose Predictor.} For predicting the 3D dose distribution given the RT plan and CT scan, we utilized a convolutional gated recurrent unit neural network, following~\cite{Witte_Sonke_2024}. We modified the architecture to be fully differentiable in our pipeline, enabling gradient-based optimization during training, while keeping it frozen during the physics-guided stage of training.

The network was trained on 350 cancer patients across multiple tumor sites, including prostate cancer, using treatment plans generated by Monaco~\cite{Elekta_Monaco}. Using a gamma pass rate criterion~\cite{Wendling2007} of 2\% and 2 mm for voxels receiving at least 10\% of the maximum dose, the model achieves a 99.6\% pass rate.

\subsection{Two-Stage Physics-Guided Training}

\textbf{First Training Stage.} During the first stage of training, the \emph{Deep RT Planner} takes as input the CT scan and the binary masks of the RT Structure Set. The network encodes these inputs, with the MLC aperture decoder predicting the binary MLC apertures and the MU decoder predicting the MU values for the complete 2-arc plan. The network is supervised using the ground truth plan and minimizes the following loss function:

\[
\mathcal{L} = \mathcal{L}_{\text{BCE}}(M_{\text{pred}}, M_{\text{true}}) + \lambda \cdot \| \text{MU}_{\text{pred}} - \text{MU}_{\text{true}} \|_1
\]

where the first term represents the Binary Cross-Entropy (BCE) loss~\cite{goodfellow2016deep} for the predicted MLC aperture \( M_{\text{pred}} \) and its ground truth \( M_{\text{true}} \). The second term is the \( L_1 \) loss for the predicted monitor units \( \text{MU}_{\text{pred}} \) compared to the ground truth \( \text{MU}_{\text{true}} \). The parameter \( \lambda \) balances the two loss components and is set to 100.

\textbf{Second Training Stage.} A key limitation of direct supervision in the first stage is the non-uniqueness of optimal solutions, as multiple MLC and MU configurations can yield clinically equivalent dose distributions, making a single ground truth plan ambiguous. While our second training stage—incorporating a dose-based loss term—does not fully resolve this ambiguity, it mitigates multi-arc redundancy by guiding the network toward a consistent dose representation. Furthermore, dose supervision evaluates the network’s output in a clinically relevant domain, aligning the optimization process with actual treatment objectives.

\begin{figure*}[t]
    \centering
    \includegraphics[width=\textwidth]{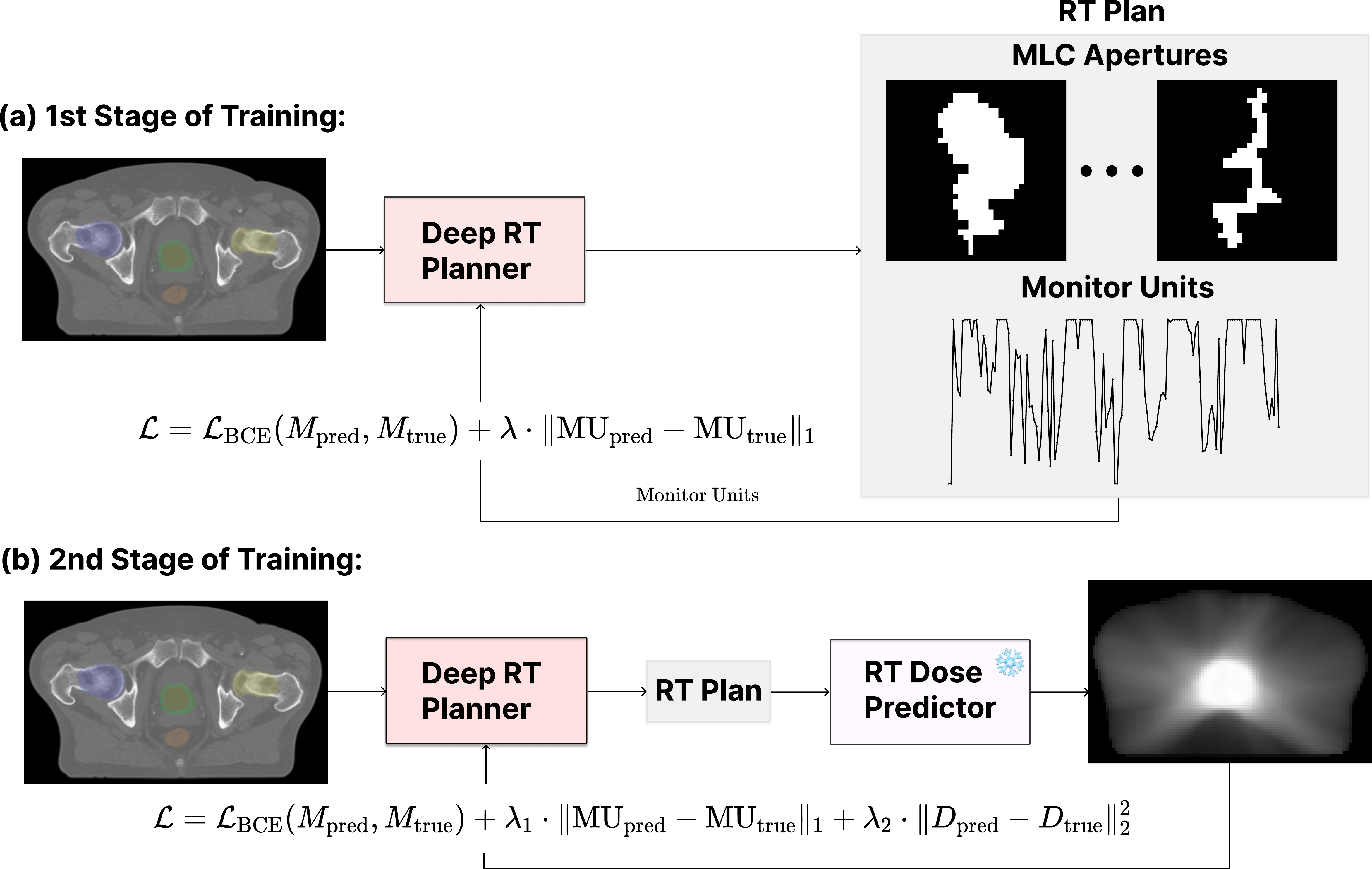}
    \caption{Two-stage physics-guided training framework: (a) The \emph{Deep RT Planner} is trained with ground truth RT plans. (b) Dose supervision via the \emph{RT Dose Predictor}.}
    \label{fig:main}
\end{figure*}

As shown in Fig.~\ref{fig:main}, the RT plan predicted by the \emph{Deep RT Planner} serves as input to the \emph{RT Dose Predictor} in a cascaded manner. This process remains fully end-to-end differentiable, allowing dose supervision to backpropagate through the pipeline and optimize the \emph{Deep RT Planner} accordingly. The second-stage loss function is defined as:

\[
\mathcal{L} = \mathcal{L}_{\text{BCE}}(M_{\text{pred}}, M_{\text{true}}) 
+ \lambda_1 \cdot \| \text{MU}_{\text{pred}} - \text{MU}_{\text{true}} \|_1
+ \lambda_2 \cdot \| D_{\text{pred}} - D_{\text{true}} \|_2^2
\]

where \(D_{\text{pred}}\) and \(D_{\text{true}}\) represent the predicted and ground truth 3D dose distributions, respectively. The parameters \(\lambda_1\) and \(\lambda_2\) control the relative weighting of the MU and dose terms and were set to 100 and 10, respectively.  

The first training stage ran for approximately 400 epochs, while the second stage lasted 100 epochs, with early stopping applied if the validation loss did not improve for 10 consecutive epochs. We used the AdamW optimizer~\cite{loshchilov2019decoupledweightdecayregularization} with cosine annealing, a learning rate of \(10^{-4}\), a weight decay of \(10^{-3}\), and a batch size of 4.

\section{Results}

\subsection{Dosimetric Comparison with Clinical Plans}

We evaluated our two-stage methodology using two different neural network architectures as the \emph{Deep RT Planner}: a 3D U-Net and a UNETR. Both networks have comparable model sizes, and this comparison aims to demonstrate that the proposed second-stage physics-guided training enhances performance independently of the \emph{Deep RT Planner} architecture. Results were averaged over a test set of 13 patients. To ensure efficient evaluation, we used CUDA-accelerated preprocessing~\cite{Cook2013}, including rotation and projection operations, and performed model inference on a single NVIDIA RTX A6000 GPU in under one second per patient.

Table~\ref{tab:results} presents the differences between predicted and ground truth (GT) dose-volume histogram (DVH) characteristics for the PTV, CTV, and OARs~\cite{Washington2020}. Each value in the table represents the difference \(\text{Predicted} - \text{GT}\), where negative values indicate that the predicted DVH characteristics are lower than those of the ground truth. For PTV and CTV, optimal performance corresponds to values closest to zero, minimizing deviation from the ground truth. In contrast, for OARs, lower dose values are preferable, meaning a negative difference relative to the ground truth indicates better sparing of healthy tissue.

The second training stage significantly improved key dosimetric metrics for the PTV/CTV, including D98, D95, and V95\%. The best performance was achieved using the 3D U-Net, which yielded a mean absolute difference of 
\( D_{95\%} = 0.59 \pm 2.23 \) Gy, 
\( D_{98\%} = 0.70 \pm 2.14 \) Gy, and 
\( V_{95\%} = -0.42 \pm 1.12\% \), 
demonstrating strong alignment between the predicted and ground truth treatment objectives.

Physics-guided training reduced OAR toxicity compared to first-stage models. A detailed analysis revealed that for the rectum, one of the most radiosensitive structures, the 3D U-Net achieved greater dose reductions than the UNETR. This trend aligns with previous findings~\cite{Heilemann_2023}. For the femoral heads, the UNETR performed better on average, though both models exhibited variability in dose metrics.

\begin{figure*}[t]
    \centering
    \includegraphics[width=\textwidth]{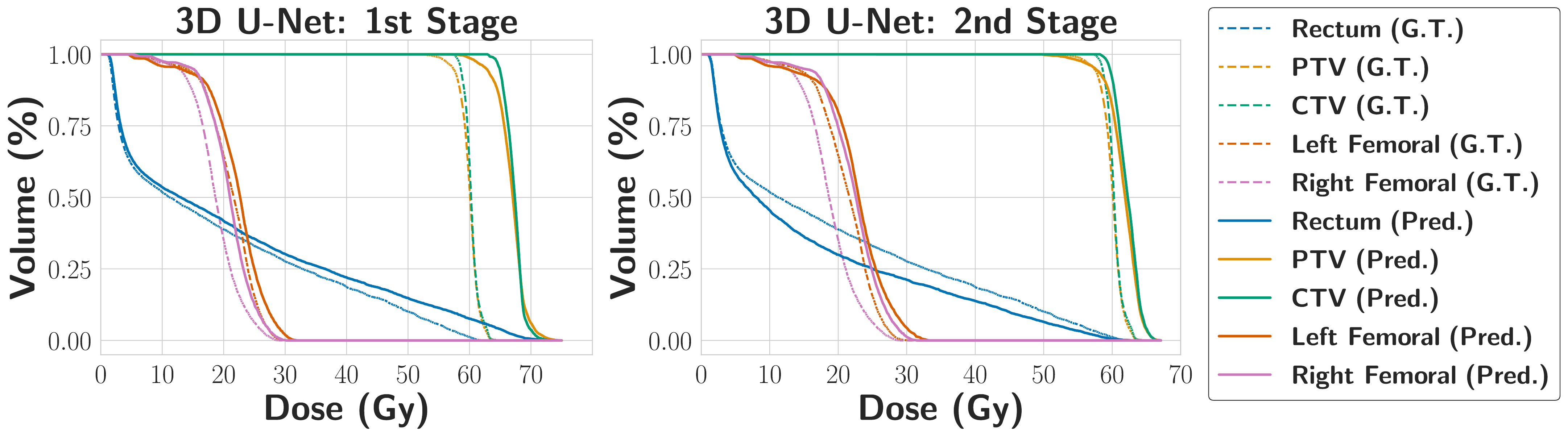}
    \caption{DVHs comparing predicted dose distributions after the 1st (left) and 2nd (right) training stages of the 3D U-Net for the same patient. Ground truth (G.T.) doses are shown as dashed lines, while predicted (Pred.) doses are shown as solid lines.}
    \label{fig:dvh}
\end{figure*}

Fig.~\ref{fig:dvh} presents DVH curves for a single patient before and after the physics-guided stage for the 3D U-Net, where we observe that the physics guidance step brings the dose objectives closer to the ground truth.

\renewcommand{\arraystretch}{1.2}
\begin{table}[h]
    \centering
    \scriptsize
    \setlength{\tabcolsep}{5pt}
    \begin{tabular}{ l l || c c | c c }
        \toprule
        & & \multicolumn{2}{c|}{\textbf{UNETR}} & \multicolumn{2}{c}{\textbf{3D U-Net}} \\
        \cmidrule(lr){3-4} \cmidrule(lr){5-6}
        ROI & Metric & 1st Stage & 2nd Stage & 1st Stage & 2nd Stage \\
        \midrule
        \multirow{3}{*}{CTV} 
        & $D_{95\%}$ (Gy) & 3.32 $\pm$ 4.34 & \textbf{-1.12 $\pm$ 2.30} & 2.94 $\pm$ 3.58 & \textbf{0.59 $\pm$ 2.23} \\
        & $D_{98\%}$ (Gy) & 2.79 $\pm$ 4.32 & \textbf{-1.30 $\pm$ 2.35} & 2.67 $\pm$ 3.59 & \textbf{0.70 $\pm$ 2.14} \\
        & $V_{95\%}$ (\%) & -1.79 $\pm$ 4.00 & \textbf{-1.68 $\pm$ 1.82} & -1.09 $\pm$ 2.41 & \textbf{-0.42 $\pm$ 1.12} \\
        \midrule
        \multirow{3}{*}{PTV} 
        & $D_{95\%}$ (Gy) & 2.59 $\pm$ 3.82 & \textbf{-1.70 $\pm$ 2.21} & 1.75 $\pm$ 3.55 & \textbf{0.42 $\pm$ 1.83} \\
        & $D_{98\%}$ (Gy) & 2.20 $\pm$ 3.96 & \textbf{-1.95 $\pm$ 2.77} & 0.92 $\pm$ 4.72 & \textbf{-0.71 $\pm$ 2.12} \\
        & $V_{95\%}$ (\%) & -1.72 $\pm$ 6.60 & \textbf{-1.63 $\pm$ 1.71} & -0.81 $\pm$ 2.55 & \textbf{-0.22 $\pm$ 1.87} \\
        \midrule
        \multirow{3}{*}{Rectum} 
        & $D_{\text{mean}}$ (Gy) & 0.48 $\pm$ 6.03 & \textbf{-0.69 $\pm$ 5.67} & 0.71 $\pm$ 5.14 & \textbf{-0.82 $\pm$ 4.45} \\
        & $D_{\text{max}}$ (Gy) & 6.08 $\pm$ 4.56 & \textbf{1.57 $\pm$ 3.16} & 5.86 $\pm$ 4.51 & \textbf{-0.07 $\pm$ 3.19} \\
        & $V_{40\text{Gy}}$ (\%) & 0.76 $\pm$ 6.49 & \textbf{-0.47 $\pm$ 5.68} & 1.08 $\pm$ 6.61 & \textbf{-1.12 $\pm$ 5.13} \\
        \midrule
        \multirow{3}{*}{Left Femoral} 
        & $D_{\text{mean}}$ (Gy) & 0.16 $\pm$ 10.22 & \textbf{-0.72 $\pm$ 6.34} & 0.35 $\pm$ 10.10 & \textbf{-0.15 $\pm$ 6.60} \\
        & $D_{\text{max}}$ (Gy) & 1.24 $\pm$ 7.65 & \textbf{-0.71 $\pm$ 5.46} & 1.19 $\pm$ 6.64 & \textbf{0.79 $\pm$ 6.21} \\
        & $V_{30\text{Gy}}$ (\%) & 0.13 $\pm$ 1.01 & \textbf{-0.71 $\pm$ 3.21} & 0.45 $\pm$ 4.57 & \textbf{-0.32 $\pm$ 3.22} \\
        \midrule
        \multirow{3}{*}{Right Femoral} 
        & $D_{\text{mean}}$ (Gy) & 1.30 $\pm$ 7.96 & \textbf{-0.73 $\pm$ 6.96} & 1.38 $\pm$ 8.03 & \textbf{-0.69 $\pm$ 5.96} \\
        & $D_{\text{max}}$ (Gy) & 2.74 $\pm$ 5.62 & \textbf{-0.09 $\pm$ 5.11} & 1.85 $\pm$ 4.90 & \textbf{1.23 $\pm$ 4.14} \\
        & $V_{30\text{Gy}}$ (\%) & 0.88 $\pm$ 2.40 & \textbf{-0.32 $\pm$ 1.15} & 0.12 $\pm$ 1.74 & \textbf{-0.21 $\pm$ 1.16} \\
        \bottomrule
    \end{tabular}
    \caption{Comparison of dose metrics for UNETR and 3D U-Net across training stages. Bold values indicate the best performance for each model. Each value represents the difference \(\text{Predicted} - \text{Ground Truth}\).}
    \label{tab:results}
\end{table}

\subsection{Gamma Pass Rate Analysis}

To further evaluate the predicted radiotherapy plans, we computed the gamma pass rate~\cite{Wendling2007} using \(3\%/3\) mm criteria for dose values exceeding 10\% of the maximum to assess overall dose distribution and for values above 90\% to focus on high-dose regions critical for target coverage.

Table~\ref{tab:gamma_pass_rates} shows that two-stage training consistently improved gamma pass rates, with the largest gain in the high-dose region, nearly doubling performance, and the 3D U-Net model achieved \(90.5\% \pm 7.3\%\).

\begin{figure*}[t]
    \centering
    \includegraphics[width=\textwidth]{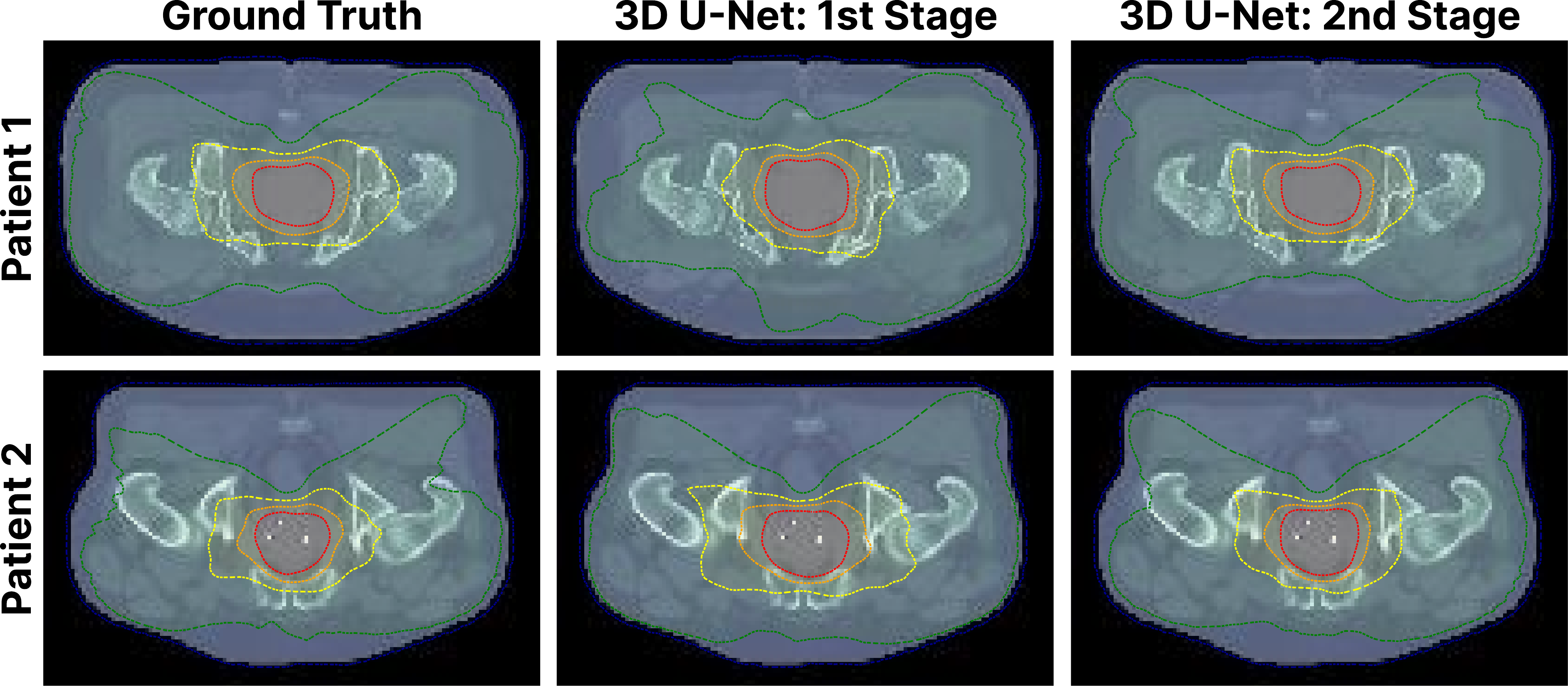}
    \caption{Isodose distributions (25\%, 50\%, 75\%, 90\% of max dose) for two patients. The ground truth (left) is compared to predictions from the 1st (middle) and 2nd (right) training stages of the 3D U-Net.}
    \label{isodose}
\end{figure*}

Fig.~\ref{isodose} compares isodose curves for two patients, showing the ground truth alongside predictions from the 1st and 2nd training stages of the 3D U-Net.

\renewcommand{\arraystretch}{1.2}
\begin{table}[ht]
    \centering
    \scriptsize
    \setlength{\tabcolsep}{5pt}
    \begin{tabular}{l || c c | c c}
        \toprule
        \multirow{2}{*}{Dose Threshold} & \multicolumn{2}{c|}{\textbf{UNETR}} & \multicolumn{2}{c}{\textbf{3D U-Net}} \\
        \cmidrule(lr){2-3} \cmidrule(lr){4-5}
        & 1st Stage & 2nd Stage & 1st Stage & 2nd Stage \\
        \midrule
        10\%  & 82.85 $\pm$ 7.30  & \textbf{85.08 $\pm$ 6.37}  & 84.70 $\pm$ 5.93  & \textbf{86.78 $\pm$ 4.63}  \\
        90\%  & 41.98 $\pm$ 25.74 & \textbf{80.40 $\pm$ 10.72} & 56.46 $\pm$ 32.68 & \textbf{90.50 $\pm$ 7.28}  \\
        \bottomrule
    \end{tabular}
    \caption{Gamma pass rates (\%) for different dose thresholds across models. Bold values indicate the best performance for each architecture.}
    \label{tab:gamma_pass_rates}
\end{table}

\section{Discussion}

We developed a physics-guided training pipeline that generates treatment plans adhering to clinical DVH criteria in under one second, whereas contemporary GPU-based planning systems require minutes~\cite{liu2021acceleratingradiationtherapydose}. Our approach directly addresses a key bottleneck in ART: time constraints.

In the first training stage, supervised on ground truth RT plans following approaches similar to those proposed in the literature~\cite{Vandewinckele_2023}, the generated plans exhibited deviations from clinical goals. The physics-guided stage introduced a clinically relevant training objective, mitigating multi-arc redundancy and guiding the network toward a consistent dose representation. As a result, PTV and CTV dose predictions aligned more closely with the ground truth, while OAR doses remained slightly below or marginally above it, demonstrating the clinical feasibility of our approach.

Regarding the \emph{Deep RT Planner}, the 3D U-Net outperformed UNETR, likely due to the latter’s need for larger training datasets~\cite{Dosovitskiy_2021}. Both architectures had similar parameter counts, but the relatively shallow transformer may have limited UNETR’s generalization.

The gamma pass rate at the 10\% dose threshold showed minimal differences between the two physics-based models, possibly explaining the variability in DVH metrics for femoral heads and the UNETR’s slight advantage in this specific metric. Conversely, the superior gamma pass rate in the high-dose region supports the observed improvement in PTV/CTV DVH metrics, underscoring the value of dose-aware supervision.

Scalability remains a challenge for broader clinical adoption. With more diverse training data, this approach could generalize to an adaptive RT agent for multiple cancer sites and treatment stages. Additionally, integrating DVH-aware loss functions~\cite{Jhanwar_2022} or leveraging geometry-aware architectures~\cite{bronstein2021geometricdeeplearninggrids} may help address these challenges.

    

\begin{credits}

\subsubsection{\ackname} The authors acknowledge the Research High-Performance Computing (RHPC) facility of the Netherlands Cancer Institute (NKI) for providing computational resources. We also extend our appreciation to Marnix Witte and Geert Wortel for their invaluable support.

\subsubsection{\discintname} This work was partially funded by Elekta Oncology Systems AB and the RVO public-private partnership grant (PPS2102) and was supported by an institutional grant from the Dutch Cancer Society and the Dutch Ministry of Health.

\end{credits}

%
%
%
\bibliographystyle{splncs04}

%
\end{document}